# Point estimation for adaptive trial designs II: practical considerations and guidance

**David S. Robertson**[1*], **Babak Choodari-Oskooei**[2], **Munya Dimairo**[3], **Laura Flight**[3], **Philip Pallmann**[4], **Thomas Jaki**[1,5]

[1] *MRC Biostatistics Unit, University of Cambridge*
[2] *MRC Clinical Trials Unit at UCL*
[3] *School of Health and Related Research (ScHARR), University of Sheffield*
[4] *Centre for Trials Research, Cardiff University*
[5] *University of Regensburg, Germany*
[*] *Corresponding author: david.robertson@mrc-bsu.cam.ac.uk*

## Abstract

In adaptive clinical trials, the conventional end-of-trial point estimate of a treatment effect is prone to bias, that is, a systematic tendency to deviate from its true value. As stated in recent FDA guidance on adaptive designs, it is desirable to report estimates of treatment effects that reduce or remove this bias. However, it may be unclear which of the available estimators are preferable, and their use remains rare in practice. This paper is the second in a two-part series that studies the issue of bias in point estimation for adaptive trials. Part I provided a methodological review of approaches to remove or reduce the potential bias in point estimation for adaptive designs. In part II, we discuss how bias can affect standard estimators and assess the negative impact this can have. We review current practice for reporting point estimates and illustrate the computation of different estimators using a real adaptive trial example (including code), which we use as a basis for a simulation study. We show that while on average the values of these estimators can be similar, for a particular trial realisation they can give noticeably different values for the estimated treatment effect. Finally, we propose guidelines for researchers around the choice of estimators and the reporting of estimates following an adaptive design. The issue of bias should be considered throughout the whole lifecycle of an adaptive design, with the estimation strategy pre-specified in the statistical analysis plan. When available, unbiased or bias-reduced estimates are to be preferred.

# 1. Introduction

Traditional clinical trials follow a design that is fixed in advance, with the data only analysed after completion when all trial participants have been recruited and have accrued outcome data[1]. In contrast, adaptive designs allow for pre-planned modifications to the trial's course based on data accumulating within the trial[2–4]. Adding controlled flexibility to the trial design in this way, while still maintaining scientific rigour, can lead to advantages in terms of efficiency and ethics compared with a traditional fixed design[2,5]. The uptake of adaptive designs in practice is becoming increasingly common[6–8], with the COVID-19 pandemic only accelerating their use[9,10]. We start by briefly highlighting some well-established types of trial adaptations below, which we will return to in this paper before focusing attention to the question of treatment effect estimation at trial completion.

*Early trial stopping: Group sequential designs*
When monitoring the accumulating outcome data of a clinical trial, it can be beneficial to have the option of stopping the trial early for safety, futility (i.e., lack of benefit), or efficacy as soon as sufficient evidence is reached to make a reliable conclusion. As continuous monitoring after every trial participant is often impractical, it is more feasible to monitor the data at (typically pre-specified) periodic intervals after a group of trial participants have accrued outcome data. This is known as a group sequential design, and at each interim analysis, the trial can be stopped early for futility or efficacy based on pre-defined stopping rules or boundaries[11]. Researchers can derive their own stopping boundaries or use standard ones such as the O'Brien-Fleming (OBF)[12] and Haybittle-Peto (HP)[13] stopping boundaries. Optimal stopping boundaries using various optimisation criteria have also been proposed[14,15].

*Treatment selection: Multi-arm multi-stage (MAMS) designs*
In some therapeutic areas, there may be several treatments or combinations of treatments awaiting evaluation in controlled clinical trials. One way of doing so efficiently is to use a MAMS design, where multiple experimental treatment arms are compared simultaneously against a single common control[16]. Similarly to group sequential designs, MAMS designs allow for early stopping of recruitment to a treatment arm for efficacy or futility (which also allows early stopping of the whole trial). These pre-planned stopping rules can be chosen to find the single best treatment[17] or all promising treatments[18] to carry forward for further evaluation. More flexible stopping rules and adaptation methods also exist[19–21]. A variant of MAMS is the drop-the-loser design[22–24], where a pre-determined number of experimental treatment arms (i.e., the worst performing ones) are dropped at each stage, typically leaving a single treatment at the final analysis.

*Population selection: Adaptive enrichment designs*

There can sometimes be large uncertainty regarding which patients would benefit from a study treatment, combined with some information to suggest that patients with certain characteristics may benefit more than others. For example, in oncology, there is a recognition that tumours can have potentially large biological heterogeneity. This motivates characterising and selecting sub-populations that are more likely to benefit from an experimental treatment[25]. Adaptive enrichment designs use interim analyses to decide which of the subpopulations should be recruited from for the remainder of the trial, where the subpopulations can be defined using biomarkers for example. Such designs can increase recruitment to the subpopulations that are estimated to receive the greatest benefit, and decrease (or stop) recruitment to the sub-populations that do not. A variety of decision rules have been proposed to select subpopulations in this way, from both Bayesian[25–27] and frequentist[28–30] perspectives.

*Changing randomisation probabilities: Response-adaptive randomisation (RAR)*

Traditional clinical trials use fixed randomisation schemes, which do not change as a result of patients' response to treatment. Alternatively, the accruing response data can be used to change the randomisation probabilities for allocating patients to treatment arms, which is known as response-adaptive randomisation (RAR). A common motivation for doing so is to allocate more patients to a treatment that is estimated to be more effective during the trial, but RAR can also be used to target other experimental objectives such as increasing the power of a specific treatment comparison[31]. Many different types of RAR procedures have been proposed for various trial contexts[31–33]. RAR can also be applied in adaptive trials in combination with other adaptations such as treatment and (sub-)population selection[34,35].

*Changing sample sizes: Sample size re-estimation*

When calculating the sample size for a trial, there may be substantial uncertainty around key design parameters (e.g., the variance of the outcome). Sample size re-estimation (or re-assessment or re-calculation) designs aim to help ensure that the sample size for a trial is appropriate, by estimating design parameters at an interim analysis and using these to re-calculate the sample size based on, for example, conditional power considerations[36–38]. This may be done in either a blinded or unblinded manner[37,39]. Sample size re-estimation can also be used in conjunction with other types of trial adaptations, such as group sequential designs[40].

Further educational material on all of these adaptive designs can be found in Burnett et al.[5], Pallmann et al.[2], and the PANDA online resource (https://panda.shef.ac.uk/).

Regardless of the type of adaptive design, it is crucial that the integrity and validity of the trial is maintained[41]. Appropriate estimation of treatment effects is a key part of trial validity, as stated in the FDA guidance on adaptive designs[42(p8)]: "It is important that clinical trials produce sufficiently reliable treatment effect estimates to facilitate an evaluation of benefit-risk and to appropriately label new drugs (*sic* treatments), enabling the practice of evidence-based medicine". The issue with estimation after an adaptive clinical trial is that the conventional end-of-trial estimates can be prone to bias, which is defined as "a systematic tendency for the estimate of treatment effect to deviate from its true value"[42(p3)]. It is clear that (all else being equal, such as the variance) it is desirable to obtain unbiased point estimates of treatment effects in order to make reliable conclusions about the treatments in a trial. While equally important, the construction of related quantities for inference, such as confidence intervals or regions, is beyond the scope of this paper so we signpost the interested reader to related literature[43,44]

This paper is the second in a two-part series that studies the issue of potential bias in point estimation for adaptive designs. In part I, we reviewed and compared current methods for unbiased and bias-reduced estimation of treatment effects after an adaptive clinical trial and critically discussed different approaches. In the current paper (part II), we consider point estimation for adaptive designs from a practical perspective, and propose a set of guidelines for researchers around the choice of estimators and the reporting of estimates following an adaptive design. We first describe the problem of estimation bias in an adaptive design in more detail in Section 2, and review current practice in Section 3. We then provide an exemplary case study in Section 4 for a real adaptive trial, using different types of unbiased and bias-reduced estimators (with R code provided) as described in part I of this paper series. We also include a simulation study and graphical representation of the sampling distribution of these different estimators. We conclude with guidance for researchers and discussion in Sections 5 and 6.

## 2. The problem of estimation bias in adaptive designs

The problem with using conventional estimators (i.e., maximum likelihood estimators, MLEs) after an adaptive trial design is that these are prone to bias. This can be because of population or treatment selection that takes place following an interim analysis (see Bauer et al.[45] for a detailed explanation of why selection process results in bias) or other types of adaptations, such as early stopping, that affect the sampling distribution of the estimator. The usual MLE is sometimes referred to as the 'naive' estimator for the trial as it does not take into account the planned and realised trial adaptations.

As introduced in part I of this paper series, different definitions of an unbiased estimator are relevant in our context, which we recapitulate below. We denote the population parameter of interest, the treatment effect, by $\theta$ and an estimator thereof by $\hat{\theta}$.

*Mean-unbiased estimators*

An estimator $\hat{\theta}$ is called *mean-unbiased* if its expected value is the same as the true value of the parameter of interest, that is, $E(\hat{\theta}) = \theta$.

*Median-unbiased estimators*

An estimator $\hat{\theta}$ is called *median-unbiased* if $P(\hat{\theta} < \theta) = P(\hat{\theta} > \theta)$, that is, if the probability of overestimation is the same as the probability of underestimation.

*Conditionally and unconditionally unbiased estimators*

An estimator is unconditionally unbiased (also known as marginally unbiased) if it is unbiased when averaged across all possible realisations of an adaptive trial. In contrast, an estimator is conditionally unbiased if it is unbiased only conditional on the occurrence of a subset of trial realisations. For example, in a drop-the-loser trial the interest will typically be on estimating the performance of the ultimately selected arm, motivating the use of a conditionally unbiased estimator (conditional on that arm being selected).

## 2.1 The potential negative impacts of reporting biased estimates

Adaptive designs play an important role in health care decision-making, increasingly providing evidence of the clinical effectiveness of an intervention as well as key secondary outcomes such as health economic ones. Failing to account for biases in point estimates can result in incorrect decision-making, potentially wasting limited resources, preventing patients from receiving effective treatments, or exposing patients to unnecessary risks. In the following subsections, we consider some potential negative impacts of reporting biased estimates.

### 2.1.1 Reporting biased primary outcomes

As highlighted by Dimairo et al.[3,4], the goal of clinical trials should be to provide reliable estimates of the treatment effect to inform accurate decision-making, but this can be compromised when an adaptive design is analysed with inappropriate statistical methods. Clearly, reporting substantially biased estimates for a primary outcome measure following an adaptive design can result in poor decisions. However, other negative impacts include the results of adaptive designs being viewed with scepticism amongst stakeholders who are aware of potential biases but do not feel they have been adequately addressed by researchers[46]. This could impede the uptake of results from an adaptive trial design or discourage research teams from using these designs in practice.

A further concern is the potential for over- or underestimation of treatment effects to affect further research. In a phase II trial, for example, ineffective treatments with exaggerated effects may be wrongly selected for further investigations in phase III trials or potentially effective treatments may not be pursued further when their effects are underestimated[47–51]. However, Wang et al.[52] and Goodman[53] suggest that group sequential trials that stop early do not produce materially biased estimates.

The consequences following biased estimates from a phase III trial could include treatments being made available to patients with an overstated benefit or treatments not being recommended for use in practice because of an understated benefit, see Briel et al.[47] for examples. Both scenarios can have a detrimental impact on patients, especially in resource limited healthcare systems such as the National Health Service (NHS) in the UK, where resources spent on a treatment with overstated benefit removes funding for alternative treatments elsewhere in the system. Mueller et al.[54] also argue that there are serious ethical problems when trialists fail to account for bias in estimates following an adaptive design, as this may violate the scientific validity of the research and social value when these estimates are used to inform clinical decision-making.

### 2.1.2 Secondary clinical outcomes

Clinical trials often collect information about a number of key secondary outcomes that may also require adjustment in an adaptive design. If these secondary outcomes are strongly correlated with the primary outcome used to inform the adaptations to the trial they will also be vulnerable to bias[55,56]. This is highlighted in the FDA guidance on adaptive designs[42], which states "It is widely understood that multiple analyses of the primary endpoint can [...] lead to biased estimation of treatment effects on that endpoint. Less well appreciated, however, is that [...] biased estimation can also apply to any endpoint correlated with the primary endpoint".

As highlighted in the benefit-risk analysis literature, there can often be a trade-off between different outcomes when developing and evaluating an intervention[57,58]. In an example reported by Briel et al.[47], a trial assessing the effectiveness of vitamin E in premature newborns was stopped early based on an interim analysis of approximately half of the total number of participants planned at the start of the trial. This early analysis showed a reduction in intracranial hemorrhage[59]. However, a later evidence synthesis showed that this trial failed to identify that vitamin E increases the risk of sepsis[60]. Failing to accurately estimate treatment effects on key secondary endpoints could result in an intervention being adopted whose safety is overestimated or whose side-effects are underestimated.

### 2.1.3 Meta-analysis and evidence synthesis literature

Meta-analysis and evidence synthesis provide frameworks for combining the results from several studies[61] and are useful tools for providing an overall understanding of what the synthesised research has found[62]. In a review of 143 trials using adaptive designs that stopped early, Montori et al.[63] found that few evidence syntheses and meta-analyses that included these trials considered the possible biases resulting from using these designs.

Several authors have argued that failing to account for adaptive designs in a meta-analysis or evidence synthesis can introduce bias[64,65,48,49]. Cameron et al.[66] explored the impact of adaptive designs in a network meta-analysis. The authors considered three alternative methods to convert outcome data derived from an adaptive design to non-adaptive designs and found that failing to account for different study designs could bias estimates of effect in a network meta-analysis. Additionally, Walter et al.[51] suggest that the estimate of treatment benefit can be calculated more accurately by applying weights to subgroups of studies with and without stopping rules.

However, there are several authors that suggest the biases are minimal[52,67–70] including Schou et al.[71] who argue that removing truncated trials from a meta-analysis leads to substantial bias, whereas including these trials does not introduce large biases. The authors therefore recommend that all studies regardless of whether they stop early should be included in meta-analyses. Finally, Marschner et al.[72] and Luchini et al.[73] provide guidance on how sensitivity analyses may be conducted to explore the impact of trials that stopped early for benefit in a meta-analysis in line with CONSORT and GRADE[74] reporting checklists.

### 2.1.4 Health economics

Increasingly clinical trials are designed with health economic objectives in mind, so that related outcomes are collected to inform a health economic analysis following the trial[75]. This may include clinical data on primary and secondary outcomes to inform parameters in a health economic model or costs and quality of life data collected directly from participants during the trial[76].

Marschner and Schou[77] discuss the underestimation of treatment effects in sequential clinical trials when they do not stop early for benefit. The authors highlight the importance of an unbiased estimate of the treatment effect for cost-effectiveness analyses using a reanalysis of the GUSTO study[78,79]. They show that the treatment effect may have been underestimated and the experimental therapy appeared less cost-effective than it actually was. Flight[80] showed, using a simulation study, that when there are high levels of correlation between primary and health economic outcomes collected during a group sequential design, bias is introduced in the point estimates (and confidence intervals) of health economic outcomes. The levels of bias may be reduced in a model-based health economic analysis but this will depend on several factors such as the data structure, correlation, and adaptive design used.

A review by Flight et al.[81] found no health economic analyses were adjusted following a group-sequential design. This potentially compromises decision-making if a decision to fund a treatment is based on biased estimates of cost-effectiveness. Additionally, patients may be penalised when a treatment is not funded based on an underestimate of cost-effectiveness, or resources may be wasted based on an overestimate of cost-effectiveness. Flight[80] extended the bias-adjusted maximum likelihood (ML) estimate approach proposed by Whitehead[82] to health economic outcomes and illustrated how this can reduce bias in a health economic analysis following an adaptive design.

## 2.2 The magnitude of the problem

In this subsection, we discuss the extent of the bias in point estimates of treatment effects as a result of interim monitoring or data-dependent stopping of an adaptive design. This might be due to the pre-specified treatment selection criteria or other stopping rules, i.e. lack-of-benefit (futility) and/or efficacy boundaries. More generally, a number of authors have discussed how the correlation between the MLE and the random design features in an adaptive design leads to bias in the MLE[83–85]. The random design features are features of the design (e.g., the number of treatment arms, sample size, allocation ratio) that are determined by the accumulating trial data and are therefore considered a random variable. As an example that we expand on below, in a two-stage group sequential design the final sample size $N$ is a random variable that is equal to $N_1$ if the trial stops at stage 1 or $N_2$ if the trial stops at stage 2. If $N = N_1$ then the MLE tends to be larger than the true treatment effect, whereas if $N = N_2$ then the MLE tends to be smaller than the true treatment effect (see Figure 1). This may be interpreted as correlation between the MLE and the random design, which leads to conditional bias in the MLE.

Indeed, it has previously been shown that the average treatment effect from a group sequential design is conditionally biased when the trial terminates early[50,51,86]. This conditional bias generally tends to be larger the 'earlier' the selection happens, that is, when the decision to stop the treatment arm or continue to the next stage is based on a relatively small amount of information. By information, we mean statistical information which is driven by the number of participants in trials with continuous and binary outcomes, and the number of primary outcome events in trials with time-to-event outcomes. However, the degree of any potential bias will depend on the stopping rules, i.e. how likely it is to stop the trial early, as well as the underlying treatment effect, as we now illustrate using results from Walter et al.[51].

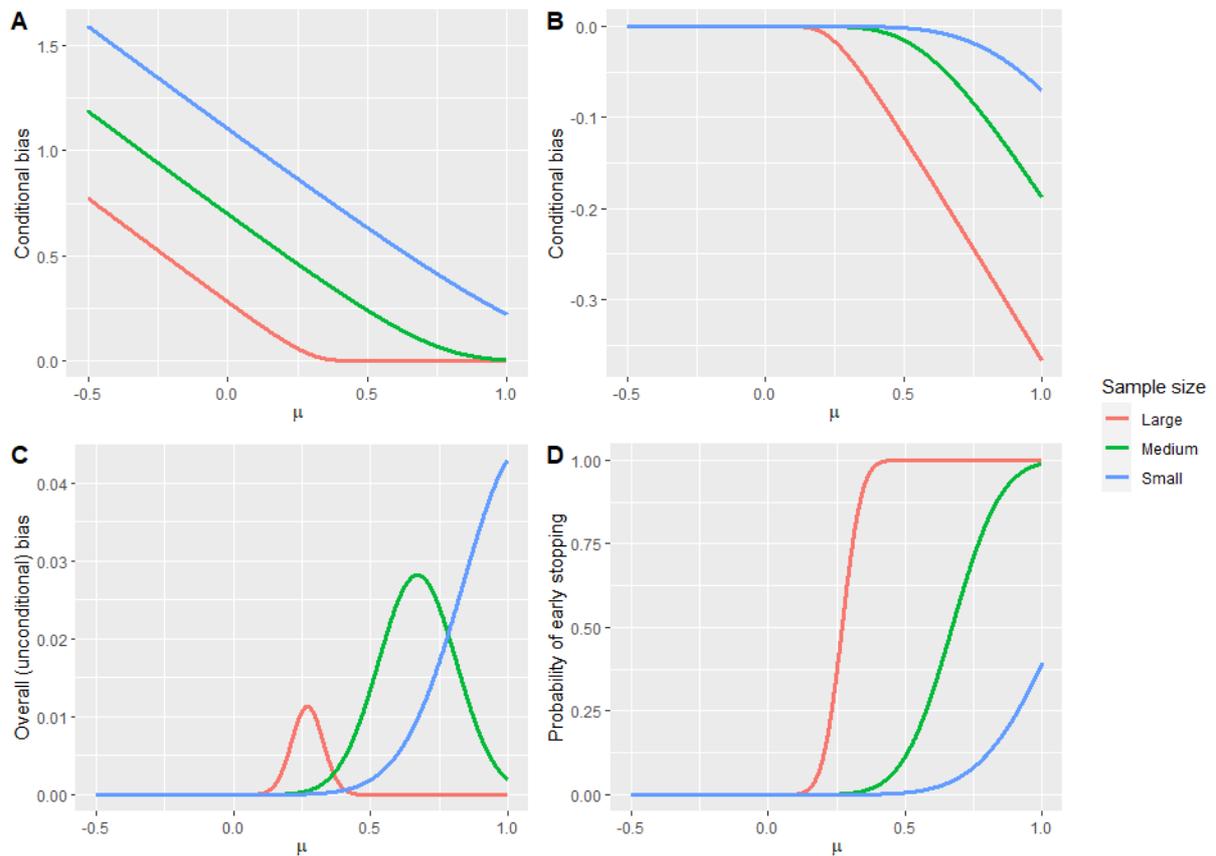

*Figure 1: Bias of the average treatment effect and probability of early stopping for a two-stage group sequential design using one-sided O'Brien-Fleming efficacy stopping boundaries under different sample sizes (small, large, medium), with overall α = 0.05. The interim analysis p-value threshold for efficacy is 0.0088. Panel A shows the expected over-estimation in trial realisations that stop early for overwhelming efficacy (i.e., conditional bias), Panel B shows the expected under-estimation in trial realisations that do not stop early for overwhelming efficacy (i.e., conditional bias at the final analysis), Panel C shows the overall (unconditional) bias and Panel D shows the probability of early stopping.*

Consider a two-stage group sequential design using one-sided OBF efficacy stopping boundaries (see Section 4 for a formal definition), where the interim analysis is conducted when 50% of the sample have provided outcome data (i.e., 50% information fraction). We assume that the study outcome variable is normally distributed with known standard deviation equal to one. Figure 1 shows the bias of the estimated treatment effect and the probability of early stopping as the true treatment effect $\mu$ varies from -0.5 to 1. The lines labelled 'large', 'medium' and 'small' correspond to studies with sample sizes of $N$ = 620, 100, and 40, respectively (which give 80% power when $\alpha$ = 0.05 to detect treatment effects of size $\mu$ = 0.2, 0.5, and 0.8).

Panel A shows that there can be a substantial positive conditional bias in the estimated treatment effect for trials that stop early for efficacy, with the magnitude of this bias increasing as the sample size decreases. Conversely, Panel B shows that there is a smaller negative conditional bias in the estimated treatment effect for trials that continue to the second stage (i.e., do not stop early), with the magnitude of this bias increasing as either the trial sample size or true treatment effect increases. Panel C shows the overall (i.e., unconditional) bias in the estimated treatment effect across all trials, which is small but positive and particularly noticeable for small sample sizes.

These results need to also be interpreted in light of Panel D, which shows that there is a very small chance of stopping for efficacy for small treatment effects when the sample size of the trial is itself small or medium. Hence, although there is a large positive conditional bias for such trials that stop early for efficacy (Panel A), and these results may then be used for recommending the adoption of a new treatment, such an event is very unlikely (Panel D). Similarly, although the negative conditional bias in trials that do not stop early can be substantial for large treatment effects (Panel B), the probability of such an event is negligible for large sample sizes (Panel D).

Previous empirical studies[87] showed that in designs with lack-of-benefit stopping boundaries the size of the selection bias for trials that reach the final stage is generally small. In fact, the bias is negligible if the experimental arm is truly effective. For trials that stopped early for lack of benefit, by definition a claim that the study treatment is better than the control is not made. Therefore, the fact that the treatment effect estimate is biased may be of less importance though results are useful in evidence synthesis. Furthermore, in designs that utilise an intermediate outcome for treatment selection it has been shown that this reduces the selection bias in the estimates of treatment effects in both selected and dropped treatment arms. In these designs, the degree of bias depends on the correlation between the intermediate and definitive outcome measures and this bias is markedly reduced by further patient follow-up and reanalysis at the planned 'end' of the trial[87].

In drop-the-loser designs, where treatment selection is done based on the relative performance (e.g., ranking) of research arms, the average treatment effect will be overestimated in the treatment arms that continue to the next stage, and will be underestimated in deselected treatment arms. In these designs, the degree of bias depends strongly on the true underlying treatment effects of the research arms, which is always unknown, the timing of treatment selection as well as the number of treatment arms to select from. Generally speaking though, in many scenarios there is a fundamental dilemma in that "the better the selection, the worse the bias"[45].

It has been shown that in drop-the-loser designs, the bias tends to be largest, and confidence intervals have incorrect coverage, where the underlying treatment effects of the treatment arms are equal, for example when all arms are under the global null or global alternative hypothesis[24,88]. More generally, bias will be smaller where the underlying effects differ amongst the treatment arms than when they are similar. In contrast, when one treatment arm has a distinctly larger treatment effect than those it is competing against, bias in the final average treatment effect is minimal for the arm which is performing best when the selection takes place. Moreover, the number of treatment arms to select from was also found to increase the degree of bias in the average treatment effects in pick-the-winner designs[89](a MAMS variant). Finally, early stopping rules that are binding increase bias compared to a design with no early stopping rules, particularly under the global null hypothesis or where only one treatment arm had the target treatment effect[45].

# 3. Review of current practice

Given that there are a variety of potential negative impacts of reporting biased estimates following an adaptive design, we reviewed the relevant literature to understand how often methods for reducing or removing bias in point estimates (as described in part I of this paper series) are used in practice. We focused on results reported from adaptive trials using the different types of adaptations described in Section 1.

## 3.1 Search strategy

A review of group sequential trials was based on pre-existing reviews known to the authors by the 30th May 2022. For other adaptive designs, we systematically searched the MEDLINE database via PubMed (from 1st Jan 2000 to 30th May 2022) using the following search terms:

| Type of adaptive design | Search terms |
| --- | --- |
| MAMS designs | ((“multi-stage”) OR (“multi stage”) OR (“multi-arm multi-stage”) OR (“multi arm multi stage”) OR (two-stage) OR (“two stage”) OR (“pick the winner”) OR (“pick-the-loser”) OR (“drop the loser”) OR (“drop-the-loser”) OR (“dose selection”) OR (“seamless”)) |
| RAR designs | ((“response adaptive”) OR (“response-adaptive”) OR (“adaptive randomisation”) OR (“adaptive randomization”) OR (“outcome adaptive”) OR (“outcome-adaptive”)) |
| Adaptive enrichment designs | ((“adaptive enrichment”) OR (“population enrichment”) OR (“patient enrichment”) OR (“enrichment design”) OR (“biomarker-adaptive”) OR (“biomarker adaptive”) OR (“subgroup selection”) OR (“subpopulation selection”) OR (“enrichment”)) |

*Table 1: Search strategy for the initial database search of MEDLINE. MAMS = multi-arm multi-stage; RAR = response-adaptive randomisation.*

## 3.2 Results

*Group sequential designs*

For group sequential designs, Stevely et al.[46] identified 68 trials that were published in leading medical journals, of which 13 (19%) were multi-arm trials. A total of 46/68 (68%) were stopped early, primarily either for efficacy (10/46, 22%) or futility (28/46, 61%). Of these trials, only 7% (3/46) disclosed the use of some form of bias correction. A subsequent review of 19 two-arm group sequential trials[90] in oncology that were stopped early found that none of these applied any bias correction to the estimated hazard ratios. Case studies also highlighted the routine lack of use of bias-corrected estimators in trials that are stopped early[91]. In summary, in trials that use group sequential designs, bias-adjusted methods are rarely used in practice and the implications are not well known[46,90–92].

*MAMS trials*

A total of 765 records were retrieved and screened for MAMS trials. Only 14/765 (1.8%) reported results. These articles were supplemented by an additional 8 trials from related work[3] within the same search period. As a result, we reviewed 22 eligible MAMS trials; phase II (n=10), phase II/III (n=9), and phase III (n=3). The vast majority of trials (95.5%, 21/22) had at least one treatment arm that was dropped early but the trial continued beyond the first interim analysis and 81.8% (18/22) used frequentist methods. Only 2/22 (9.9%) of the trials reported the use of a bias-adjusted point estimator of the treatment effect. One described the use of a bias-adjustment for the mean of the selected dose in stage 2 (which was then used to derive an adjusted t-test)[93] while the other used a uniformly minimum variance conditionally unbiased estimator[94].

*RAR designs*

There were 59 records retrieved that used a RAR design; of which 22 were randomised trials. Of these 22, 17 were reporting interim (n=4), both interim and final results (n=1) and final results (n=12); phase II (n=10), phase I/II (n=2), phase II/III (n=1), phase III (n=3) and phase IV (n=1). Of the 5 that reported interim results, 2 were stopped early for futility, 2 had a treatment that graduated (i.e. was declared successful) for further evaluation at phase III, and the remaining was stopped early for efficacy. Most of the trials (76.5%, 13/17) used Bayesian methods with the remaining using hybrid (frequentist and Bayesian, n=2) and frequentist methods (n=2). None of the trials used bias correction methods.

*Adaptive enrichment trials*

Of the 528 records screened, only 1 trial was an adaptive enrichment trial reporting results . There were an additional 2 known trials not retrieved in the search. Only 1 of these 3 trials reported interim results and was stopped early for futility. Enrichment was triggered in 1 trial. All 3 trials used frequentist methods and none used any bias correction methods.

A summary of the results of the systematic search is given in Table 2. As can be seen, across the adaptive designs considered, unbiased or bias-adjusted treatment effect estimates are currently rarely used and reported.

| Type of adaptive design | Number of records screened | Number of randomised trials that reported results | Number of randomised trials that reported an unbiased or bias-adjusted estimate |
|---|---|---|---|
| MAMS | 773 | 22 | 2 |
| RAR | 59 | 17 | 0 |
| Adaptive enrichment | 530 | 3 | 0 |

*Table 2: Summary of systematic search of adaptive designs from 1st January 2000 to 30th May 2022.*

These results are in stark contrast with some of the recommendations for best practice that have been made around bias-adjusted analyses in widely-used guidance on adaptive designs (namely, the FDA guidance[42] and the Adaptive designs CONSORT Extension[3,4], ACE). The FDA guidance stresses the importance of using "methods for adjusting estimates to reduce or remove bias associated with adaptations and to improve on performance such as the mean squared error", stating that "Such methods should be prospectively planned and used for reporting results when they are available". In particular, for group sequential designs "a variety of methods exist to compute estimates and confidence intervals that appropriately adjust for the group sequential stopping rules … To ensure the scientific and statistical credibility of trial results and facilitate important benefit-risk considerations, an approach for calculating estimates and confidence intervals that appropriately accounts for the group sequential design should be prospectively planned and used for reporting results." As for the CONSORT ACE Extension, the guidance is much less prescriptive, but nevertheless recommends the discussion of "the implications of … potential bias and imprecision of the treatment effects if naïve estimation methods were used". The relevant parts of both these guidance documents are quoted in full in Appendix A.1.

## 4. Case study: Group sequential design

In this section, we illustrate how different types of unbiased and bias-reduced estimators (as reviewed in part I of this paper series) can be used in practice for a group sequential design that uses OBF stopping boundaries, which we now briefly describe. In a group sequential design, participants are allocated to the treatments and the accumulating data are analysed after each complete group of data becomes available. When using OBF efficacy boundaries, the nominal significance levels needed to reject the null hypothesis increases as the trial progresses, i.e., it is more difficult to reject $H_0$ at earliest analyses. Given the standardised test statistics $Z_k$ for group $k = 1, ..., K$, the one-sided OBF boundaries and stopping rules take the following form[11]:

> After group $k = 1, ..., K - 1$
>     if $Z_k \geq C(K,\alpha)\sqrt{(K/k)}$      stop, reject $H_0$
>     otherwise      continue to group $k+1$
> After group $K$
>     if $Z_k \geq C(K,\alpha)$      stop, reject $H_0$
>     otherwise      stop, do not reject $H_0$

Here, the values of $C(K,\alpha)$ are chosen to ensure that the overall type I error probability for the $K$ stage trial is controlled at preset level $\alpha$.

We use a group sequential design as our case study in order to illustrate the widest range of different estimators (both unconditional and conditional). As well, some adaptive designs (e.g., certain types of MAMS designs) can be viewed as an extension of group sequential designs and therefore we can illustrate more general underlying principles of point estimation.

We consider the phase III MUSEC (MUltiple Sclerosis and Extract of Cannabis) trial, as described by Bauer et al.[95] and Zajicek et al.[96]. This is an example of a two-stage group sequential design where the trial continued to the second stage (as the criterion for early stopping was not met). The MUSEC trial investigated the effect of a standardised oral cannabis extract (CE) on muscle stiffness for adult patients with stable multiple sclerosis. The primary endpoint was a binary outcome - whether or not a patient had 'relief from muscle stiffness' after 12 weeks of treatment, based on a dichotomised 11 point category rating scale. A two-stage group sequential design with early stopping for superiority using the OBF boundary was used, with a pre-planned maximum total sample size of 400 subjects (200 per arm) and an unblinded interim analysis planned after 200 subjects (100 per arm) had completed the 12 week treatment course.

In the actual trial, an unblinded sample size re-estimation based on conditional power considerations[95] was also carried out at the interim analysis, which reduced the total planned sample size from 400 to 300. For the purpose of illustrating the calculation of a larger range of adjusted estimators, we ignore this sample size re-estimation in what follows. If we were to take into account the sample size re-estimation then the methods for adaptive group sequential designs would apply (see Section 6 of part I of this series), but only median-unbiased estimators are available in that setting. For more general guidelines around best practice in this context, see Section 5.

Table 3 summarises the observed data from the trial at the interim and final analyses, as well as the standardised test statistics and the OBF efficacy stopping boundaries. As can be seen, at the interim analysis the boundary for early rejection of the null hypothesis (no difference in the proportion of subjects with relief from muscle stiffness between treatment arms) was almost reached, with the standardised test statistic being close to the stopping boundary.

|  | Interim Data | | Final Data | |
| --- | --- | --- | --- | --- |
|  | Placebo | CE arm | Placebo | CE arm |
| Number of subjects with relief from muscle stiffness | 12 | 27 | 21 | 42 |
| Total number of subjects | 97 | 101 | 134 | 143 |
| Standardised test statistic | 2.540 | | 2.718 | |
| OBF boundary | 2.797 | | 1.977 | |

*Table 3: Observed data from the MUSEC trial at the interim and final analyses, with standardised test statistics and O'Brien-Fleming (OBF) group sequential boundary (one-sided, with early stopping only for superiority).*

## 4.1 Calculation of unbiased and bias-adjusted estimators

Using the observed data from the MUSEC trial, we now demonstrate how to calculate various unbiased and bias-adjusted estimators for the treatment difference, from both a conditional and unconditional perspective. More formally, letting $p_{CE}$ and $p_0$ denote the response probability for patients on CE and the placebo respectively, we consider estimators of $\theta = p_{CE} - p_0$. R code to obtain these estimators is provided in the supporting materials.

The conventional end-of-trial estimator for the treatment difference, i.e., the overall MLE, is given by $\hat{\theta} = \hat{p}_{CE} - \hat{p}_0$, where $\hat{p}_{CE}$ and $\hat{p}_0$ are the observed proportions of successes on the CE and placebo arms respectively. In what follows, it is also useful for illustrative purposes to consider the MLE calculated just using the interim data (stage 1 data), denoted $\hat{\theta}_1$, as well as the MLE calculated just using the stage 2 data (i.e., only the data from after the interim analysis), denoted $\hat{\theta}_2$. These estimators are inefficient (and potentially unethical) since they 'discard' patient data, so we are not recommending that these are used in practice.

*Unconditional perspective*

From an unconditional perspective, we want to estimate $\theta$ regardless of the stage that the trial stops, and are interested in the bias as averaged over all possible stopping times, weighted by the respective stage-wise stopping probabilities. More formally, letting $T$ be the random variable denoting the stage that the trial eventually stops, we define the unconditional bias of an estimator $\hat{\theta}$ as

$$bias(\hat{\theta}) = E_\theta[\hat{\theta}] - \theta = \sum_{k=1}^{2} E_\theta[\hat{\theta}|T=k]\Pr_\theta[T=k] - \theta$$

In the two-stage trial setting, Emerson[97] presented an analytical expression for this bias of the overall MLE, which depends on the unknown value of $\theta$:

$$bias(\hat{\theta}) = \frac{I_2 - I_1}{I_2\sqrt{I_1}}\phi(e - \theta\sqrt{I_1})$$

where $e$ denotes the efficacy stopping boundary, $I_1$ and $I_2$ denote the (observed) information at stage 1 and stage 2 respectively and $\phi$ denotes the probability density function (pdf) of a standard normal distribution. The full definitions of the information $I_1$ and $I_2$ for our trial context are given in Appendix A.2, which depend on the number of observed successes. Following Whitehead[82], we can use this expression to calculate an unconditional bias-corrected MLE $\tilde{\theta}$ (UBC-MLE), which is the solution of the equation $\tilde{\theta} = \hat{\theta}_{obs} - bias(\tilde{\theta})$, where $\hat{\theta}_{obs}$ is the observed value of the overall MLE at the final stage.

Alternatively, the uniformly minimum variance unbiased estimator (UMVUE) can be calculated by using the Rao-Blackwell technique on the stage 1 MLE $\hat{\theta}_1$, which is unconditionally unbiased (see Section 4.1 of part I of this paper series for further details). More formally, the UMVUE in our two-stage trial context is given by $E[\hat{\theta}_1 | (T=2, \hat{\theta}_{obs})]$, with the following closed-form expression:

$$UMVUE = \hat{\theta}_{obs} - \frac{\sqrt{I_2 - I_1}}{\sqrt{I_1 I_2}} \frac{\phi\left(\frac{e - Z_2\sqrt{I_1/I_2}}{\sqrt{(I_2 - I_1)/I_2}}\right)}{\Phi\left(\frac{e - Z_2\sqrt{I_1/I_2}}{\sqrt{(I_2 - I_1)/I_2}}\right)}$$

where $Z_2$ is the (observed) overall standardised test statistic at stage 2, and $\Phi$ represents the cumulative distribution function (cdf) of a standard normal distribution.

A median unbiased estimator (MUE) can also be calculated, which depends on a choice of the ordering of the sample space with respect to evidence against the null hypothesis (see Section 4.2 of part I of this paper series for further details). In what follows, we use stagewise ordering, which has desirable properties described by Jennison and Turnbull[11]. This allows the use of the *p*-value function P(θ) to find the MUE, which is the solution to the equation $P(\hat{\theta}_{MU}) = 0.5$. The formula for the *p*-value function (for a trial that continues to the second stage) is as follows:

$$P(\theta) = \int_{-\infty}^{e} \int_{z_2}^{\infty} f_2\left((x_1, x_2), (\theta\sqrt{I_1}, \theta\sqrt{I_2}), \begin{pmatrix} 1 & \sqrt{I_1/I_2} \\ \sqrt{I_1/I_2} & 1 \end{pmatrix}\right) dx_2 dx_1$$

where $f_2((x_1, x_2), \mu, \Sigma)$ is the density of a bivariate normal distribution with mean μ and covariance matrix Σ evaluated at the vector $(x_1, x_2)$. See also the R code provided in the supporting materials.

*Conditional perspective*
From a conditional perspective we are interested in estimation conditional on the trial continuing to stage 2. We define this conditional bias of an estimator $\hat{\theta}$ as $E_\theta[\hat{\theta} | T = 2] - \theta$. In the context of group sequential trials, as argued by several authors[98–102], the conditional bias of an estimator is also an important consideration: given that the study has in fact stopped with $T = 2$, we can use this knowledge in our bias calculations. As well, while the unconditionally unbiased estimators are unbiased overall, they tend to overestimate the treatment effect when there is early stopping and underestimate the effect when the trial continues to the end. The authors see value in both the conditional and unconditional perspective. As the unconditional estimators tend to be biased once the stopping reason is known they do, however, have a slight preference for conditional estimators. Nonetheless, there is no consensus in the literature and we provide a few example quotations illustrating this in Appendix A.2.

We can calculate an analytical expression for the conditional bias of the overall MLE (i.e., at the final stage), which is given below and again depends on the unknown true parameter θ:

$$conditional\ bias(\hat{\theta}) = -\frac{\sqrt{I_1}}{I_2} \frac{\phi(e - \theta\sqrt{I_1})}{\Phi(e - \theta\sqrt{I_1})}$$

Using this expression, we can calculate a conditional bias-corrected MLE $\tilde{\theta}_c$ (CBC-MLE), which is the solution of the equation $\tilde{\theta}_c = \hat{\theta} - conditional\ bias(\tilde{\theta}_c)$. As well, the uniformly minimum variance conditionally unbiased estimator (UMVCUE) can be calculated by again using the Rao-Blackwell technique, but this time applied to the stage 2 MLE $\hat{\theta}_2$ (i.e. excluding the stage 1 data), which is conditionally unbiased. See Section 4 of part I of this paper series for details. More formally, the UMVCUE is given by $E[\hat{\theta}_2 | (T = 2, \hat{\theta}_{obs})]$, resulting in a closed-form expression as follows:

$$UMVCUE = \hat{\theta}_{obs} - w_1 \frac{\phi\left(w_2(\hat{\theta}_{obs} - e/\sqrt{I_1})\right)}{\Phi\left(w_2(\hat{\theta}_{obs} - e/\sqrt{I_1})\right)}$$

where $w_1 = \dfrac{1}{(I_2 - I_1)\sqrt{I_1^{-1} + (I_2 - I_1)^{-1}}}$, $w_2 = I_1\sqrt{I_1^{-1} + (I_2 - I_1)^{-1}}$

Table 4 gives the values of all of the estimators described above, calculated using the observed data and OBF stopping boundaries from the MUSEC trial. For illustration purposes, we also calculated the standard error (SE) for all estimators using a parametric bootstrap approach assuming that the true unknown difference in proportions (θ) is equal to 0.14, which should be treated with caution as they will vary depending on this key assumption - see Section 4.2 for a simulation study that gives the values of the SEs for different assumed values of θ. R code for calculating the SEs is given in the supporting materials.

| Type of estimator | Estimator | Difference in proportions (SE) | Relative difference to overall MLE |
|---|---|---|---|
| MLE/naive | **MLE (overall)** | **0.1370 (0.054)** | - |
| Unconditionally unbiased/bias-adjusted | MLE (stage 1) | 0.1436 (0.057) | +5% |
| | Median unbiased estimator (MUE) | 0.1341 (0.054) | -2% |
| | UMVUE | 0.1278 (0.054) | -7% |
| | Bias-corrected MLE (UBC-MLE) | 0.1328 (0.055) | -3% |
| Conditionally unbiased/bias-adjusted | MLE (stage 2) | 0.1139 (0.111) | -17% |
| | UMVCUE | 0.1724 (0.071) | +26% |
| | Bias-corrected MLE (CBC-MLE) | 0.1909 (0.073) | +39% |

*Table 4: Naive, unconditionally and conditionally unbiased / bias-adjusted estimates calculated using the observed data and O'Brien-Fleming efficacy stopping boundaries from the MUSEC trial. Standard errors (SEs) are calculated using a parametric bootstrap approach with $10^5$ replicates, assuming that the true difference in proportions is equal to 0.14.*

The overall MLE is 0.1370 (with a SE of 0.054), and is the comparator for all the other estimators in Table 4 since it is the conventional end-of-trial point estimate. Starting with the unconditionally unbiased and bias-adjusted estimators, the stage 1 MLE is slightly larger (0.1436), but this is based on only the stage 1 data and hence is slightly inefficient: it has an information fraction of 0.795 and a SE of 0.057. The MUE, UBC-MLE and UMVUE are all slightly lower than the MLE, although they are all within 0.01 in absolute terms (i.e. within 7% in relative terms). This downward correction is intuitive – we would expect the MLE to overestimate the magnitude of θ averaged over the possible stopping times: if $\hat{\theta}_1$ is sufficiently larger than θ, the trial stops with $T = 1$ and the MLE equals $\hat{\theta}_1$, whereas if $\hat{\theta}_1$ is lower than θ by a similar amount, the trial can continue, allowing the stage 2 data to reduce the negative bias of the overall MLE. The SEs of these estimators are all very similar to that of the MLE under the assumption of a true difference in proportions of 0.14, reflecting the small corrections to the MLE. Finally, we see that MUE > UBC-MLE > UMVUE, which reflects the fact that the MUE is not mean-unbiased, and the UBC-MLE will also be expected to have residual mean bias as it is not exactly mean-unbiased (see Section 4.2 for simulation results).

Moving on to the conditionally unbiased and bias-adjusted estimators, the stage 2 MLE is substantially lower (0.1139) than the overall MLE (and indeed any of the other estimators conditional or unconditional). However, the information fraction for stage 2 was only 0.205 ignoring the 0.795 from stage 1, and hence this estimator has a substantially higher variability with a (conditional) SE of 0.111. Both the CBC-MLE and the UMVCUE are noticeably larger than the overall MLE (in relative terms 39% and 26% larger respectively). An upward correction is intuitive from a conditional perspective: there is downward selection pressure on the stage 1 MLE $\hat{\theta}_1$, since if $\hat{\theta}_1$ is sufficiently larger than θ then the trial does not continue to stage 2. Given that the stage 1 MLE was almost large enough for the standardised test statistic to cross the OBF stopping boundary (note that on the test statistic scale, the OBF stopping boundary at stage 1 was 0.1581), the relatively large correction to the overall MLE is not too surprising. Both estimators have substantially lower (conditional) SE than the stage 2 MLE, since they are utilising all of the trial data. However the conditional SEs are larger than the unconditional ones. This is a general property of conditional estimators: by conditioning, the information that is contained in the statistic that is conditioned on (in this case, the stopping stage) is lost. Finally, we see that CBC-MLE > UMVCUE, which again reflects the residual mean (conditional) bias in the CBC-MLE (see Section 4.2 for simulation results).

As pointed out by an anonymous reviewer, the observed stage at which the trial stops (or more generally, the observed study design) will imply something about the treatment effect which can then be taken into account. In our case study, the study proceeds to the second stage, which implies that the MLE is on average conditionally biased downwards[99] (see Panel B in Figure 1). Therefore, it arguably is undesirable to apply an unconditional adjustment that adjusts the MLE further downwards. Instead, a conditional approach may make more sense because it will adjust the MLE upwards to take account of the fact that the MLE has a conditional negative bias (see Panel B in Figure 1). These trends are all reflected in the results presented in Table 4. See also the discussion on the conditional versus unconditional perspective in Section 5.1.

In summary, for the MUSEC trial data, the use of different estimators can give noticeably different values for the estimated treatment effect, particularly when considering a conditional versus unconditional perspective. This could influence the interpretation of the trial results in certain cases, and highlights the importance of pre-specifying which estimator(s) will be reported following an adaptive design. The choice of estimator(s) will depend on what the researchers wish to achieve regarding the estimand in question. There will be pros and cons for the different estimators, one key example being the bias-variance trade-off. We explore these issues further in Section 5. We also note that there is a strong link between design and estimation - the estimated values above depend on the design of the trial, and would be different if (for example) the design had also included futility stopping boundaries.

### 4.2 Simulation study

Since the point estimates presented above represent one realisation of the trial data given the trial design, in this subsection we carry out a simulation study to investigate the performance of the estimators under different scenarios. We stress however that (unlike when calculating the standard errors of the estimates) we have *not* used the assumed unknown value for the underlying treatment effect ($\theta$) to calculate the unbiased and bias-adjusted estimates in Table 4. As can be seen from the formulae in Section 4.1, these estimators do not depend explicitly on $\theta$, but only on the observed data and efficacy stopping boundary (in this case). Hence, the point estimates presented in the third column of Table 4 would not change under different values of $\theta$.

To demonstrate the bias-variance properties of the estimators when averaged over many trial realisations following the two-stage design of the MUSEC trial, we ran simulations under different values of $\theta$ (0.10, 0.14 and 0.18). To do so, we used the (asymptotic) canonical joint distribution of the standardised test statistics at each stage[11], which is bivariate normal - for further details, see the R code included in the supporting materials. For each value of θ, we simulated $10^5$ trial replicates and calculated the mean values of the point estimators as well as their standard deviations across the trial replicates. Note that the unconditional estimators are all equal to the stage 1 MLE for the trial realisations that stop at the interim analysis. This is by definition for the overall MLE (since this is the MLE calculated at the stage the trial stops at) and the MUE, while for the UMVUE and the UBC-MLE we use the fact that the MLE at the interim analysis is unconditionally unbiased[51]. As for the conditional estimators, these are all conditional on the trial continuing to stage 2 (and so are calculated using $10^5$ trial realisations that all continue to stage 2, see the R code for further details). The simulation results are displayed in Table 5 below.

| Type of estimator | Estimator | Difference in proportions (SE) | | |
| --- | --- | --- | --- | --- |
| | | θ = 0.10 | θ = 0.14 | θ = 0.18 |
| MLE/naive | MLE (overall) | 0.103 (0.054) | 0.144 (0.054) | 0.184 (0.053) |
| Unconditionally unbiased/bias-adjusted | MLE (stage 1) | 0.100 (0.057) | 0.140 (0.057) | 0.180 (0.057) |
| | Median unbiased estimator (MUE) | 0.101 (0.053) | 0.142 (0.054) | 0.182 (0.054) |
| | UMVUE | 0.100 (0.052) | 0.140 (0.054) | 0.180 (0.055) |
| | Bias-corrected MLE (UBC-MLE) | 0.101 (0.054) | 0.142 (0.055) | 0.183 (0.054) |
| Conditionally unbiased/bias-adjusted | MLE (stage 2) | 0.100 (0.111) | 0.140 (0.111) | 0.180 (0.111) |
| | UMVCUE | 0.100 (0.062) | 0.140 (0.071) | 0.179 (0.080) |
| | Bias-corrected MLE (CBC-MLE) | 0.111 (0.067) | 0.154 (0.073) | 0.194 (0.078) |

*Table 5: Simulation results showing the mean values of the point estimators and the corresponding standard errors (SE) under different assumed values of θ. There were $10^5$ trial replicates for each value of θ. The probability of stopping at stage 1 was 0.15, 0.37 and 0.65 for θ = 0.10, 0.14 and 0.18, respectively.*

The results for the mean values of the point estimators are what we would expect: on average, the unbiased estimators are unbiased whereas the naïve and bias-adjusted estimators have a (small) positive bias due to the early stopping for efficacy. The overall MLE has the largest mean bias out of the unconditional estimators, although this is less than 0.005 in absolute terms across the different values of θ. The MUE and UBC-MLE have a residual positive bias (less than 0.003 in absolute terms), reflecting how they are not mean unbiased. Looking at the conditional estimators, the CBC-MLE has a larger positive bias (up to 0.014 in absolute terms) while having a similar SE to the UMVCUE, and so would likely not be recommended for use in this trial context.

Tables 4 and 5 demonstrate that while on average, the different point estimators will be close together, for a particular trial realisation, the estimates may be quite different. The differences between the estimators we see in Table 4 are a consequence of the observed data for the MUSEC trial, which were quite 'extreme' in the sense that the stage 1 MLE was very close to the stopping boundary and substantially larger than the stage 2 MLE.

As for the SEs, for the unconditional estimators, the stage 1 MLE has the highest SE, reflecting how it only uses the stage 1 data (with an information fraction of 0.795). The other unconditional estimators have very similar SEs, which change little as θ increases. For the conditional estimators, the stage 2 MLE has a substantially higher SE since it only uses the stage 2 data (with an information fraction of only 0.205). The UMVCUE and CBC-MLE have similar SEs; however, their SEs increase as θ increases. This reflects how the stage 1 and stage 2 data will be expected to have a larger discrepancy as θ increases, since the stage 1 MLE will have to be below the efficacy stopping boundary of 0.1581 in order for the trial to continue to the second stage.

Like in Section 2.2, it is informative to also report separate means for the unconditional estimators for the trial replicates that stop early at the interim analysis and those that continue to stage 2. Table 6 shows these mean values of the unconditional point estimators as well as their corresponding standard errors across $10^5$ trial replicates. We see that for the trial replicates that stop at the interim analysis, the stage 1 MLE has a substantial positive conditional bias, particularly for θ = 0.10 (but note that the probability of stopping in this case is only 0.15). Conversely, for the trial replicates that continue to stage 2, all of the unconditional estimators are conditionally biased, with a noticeable negative conditional bias across all three values of θ. These results need to be interpreted in light of the probability of stopping at stage 1, which was 0.15, 0.37 and 0.65 for θ = 0.10, 0.14 and 0.18, respectively. Overall, the results again demonstrate that even if estimators are unconditionally unbiased, they may have considerable conditional bias.

|  | Estimator | Difference in proportions (SE) | | |
|---|---|---|---|---|
|  |  | θ = 0.10 | θ = 0.14 | θ = 0.18 |
| **Trial stops early at the interim analysis** | MLE (stage 1) | 0.188 (0.025) | 0.197 (0.031) | 0.212 (0.038) |
| **Trial continues to stage 2** | MLE (overall) | 0.087 (0.043) | 0.113 (0.038) | 0.132 (0.033) |
|  | MLE (stage 1) | 0.084 (0.045) | 0.106 (0.038) | 0.120 (0.030) |
|  | Median unbiased estimator (MUE) | 0.086 (0.041) | 0.109 (0.034) | 0.126 (0.027) |
|  | UMVUE | 0.084 (0.039) | 0.106 (0.030) | 0.120 (0.023) |
|  | Bias-corrected MLE (UBC-MLE) | 0.085 (0.041) | 0.110 (0.036) | 0.128 (0.032) |

*Table 6: Simulation results showing the mean values of the unconditional point estimators and the corresponding standard errors (SE) under different assumed values of θ, separated by trial replicates that stop at the interim analysis and those that continue to stage 2. There were $10^5$ trial replicates in total for each value of θ. The probability of stopping at stage 1 was 0.15, 0.37 and 0.65 for θ = 0.10, 0.14 and 0.18, respectively.*

Apart from summarising the mean values and SEs, it is also useful to look at the whole sampling distribution of the point estimators. Figure 2 shows these sampling distributions, assuming θ = 0.14 and with $10^5$ trial replicates.

For the unconditional estimators (except for the stage 1 MLE), the distributions are a mixture from the trial replicates that stopped in stage 1 and those that continued to stage 2 (recall that for those trial replicates that stopped in stage 1, all the unconditional estimators are equal to the stage 1 MLE). It is interesting to note that the sampling distribution of the MUE is substantially smoother than those of all the other estimators (ignoring the stage 1 MLE), particularly compared with the UMVUE and the overall MLE. As for the conditional estimators, the stage 2 MLE has a substantially wider sampling distribution than the others, reflecting how it uses less information. Meanwhile, the sampling distributions of the UMVCUE and the CBC-MLE appear quite similar.

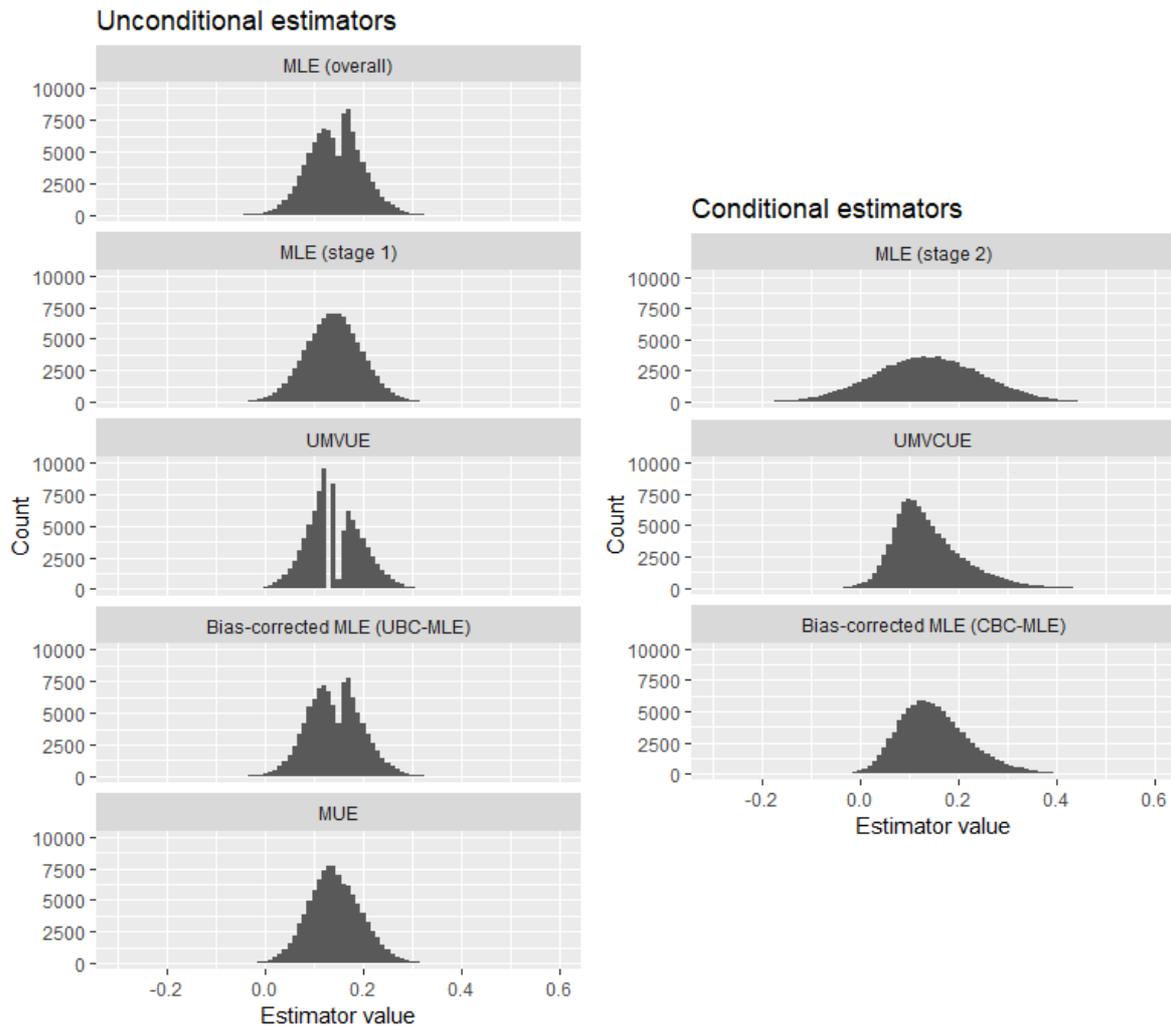

*Figure 2: Sampling distributions of the point estimates from $10^5$ trial replicates, assuming that θ = 0.14. MLE = Maximum Likelihood Estimator; MUE = Median Unbiased Estimator; UMVCUE = Uniformly Minimum Variance Conditionally Unbiased Estimator; UMVUE = Uniformly Minimum Variance Unbiased Estimator.*

# 5. Guidance: best practices for point estimation in adaptive designs

In this section, we give guidance on the choice of estimators and the reporting of estimates for adaptive designs. This builds on the relevant parts of the FDA guidance for adaptive designs[42] and the ACE[3,4]. The issue of estimation and potential bias should be considered throughout the whole lifecycle of an adaptive trial, from the planning stage to the final reporting and interpretation of the results. Indeed, the design and analysis of an adaptive trial are closely linked, and one should not be considered without the other. In what follows, our main focus is on the confirmatory setting where analyses are fully pre-specified, but some of the principles can also apply to more exploratory settings, particularly around the choice of estimators and the final reporting of trial results.

## 5.1 Planning stage

The context, aims and design of an adaptive trial should all inform the analysis strategy used, which includes the choice of estimators. These decisions should not only be left to trial statisticians, but also be discussed with other trial investigators to ensure that it is consistent with what they want to achieve. Firstly, it is necessary to decide on what exactly is to be estimated (that is, the estimands of interest - see also Appendix A.2). Secondly, the desired characteristics of potential estimators should be decided. Two key considerations are as follows:

- *Conditional versus unconditional perspective*: The choice of whether to look at the conditional or unconditional bias of an estimator will depend on the trial design. For example, in a drop-the-losers trial where only a single candidate treatment is taken forward to the final stage, a conditional perspective reflects the interest being primarily in estimating the effect of the successful candidate. On the other hand, for group sequential trials, the unconditional perspective is recognised as being an important consideration (see Appendix A.2). As seen in the simulation study in Section 4.2, it can be the case that these different point estimators are on average similar over repeated realisations of the trial, but for a single realisation are markedly different. As well, the standard errors of the conditional estimators can be larger than those for the unconditional estimators. More generally, (as pointed out by an anonymous reviewer) in situations where the observed design (see Section 2.2) implies a directionality to the MLE bias, the conditional estimation that takes this directionality into account may be preferable. Stopping a group sequential trial for efficacy at the first interim analysis (over-estimate) or at the final analysis (under-estimate) are examples of this situation.

    A new perspective on the question of conditional versus unconditional inference has recently been provided by Marschner[85]. This work presents a unifying formulation of adaptive designs and a general approach to their analysis, which is based on the partitioning of the overall unconditional information into its two component sources. More precisely, the unconditional likelihood can be expressed as the product of the 'design likelihood' (i.e., the information contained in the realised design) and the 'conditional likelihood' (conditioned on the realised design). Rather than advocating for or against unconditional inference over conditional inference in general, the framework allows for the exploration of the extent to which conditional bias is likely to be present within a given sample (using meta-analysis techniques). For further details, we refer the reader to Marschner[85].

- *Bias-variance trade-off*: Typically there will be a trade-off between the bias and variance of different estimators. Depending on the context and aims of the trial, different relative importance may be given to the two. For example, in a phase II trial where a precise estimate of the treatment effect is needed to inform a follow-up confirmatory study, the variance of an estimator may be of greater concern, whereas in a definitive phase III trial an unbiased estimate of treatment effect is key for real-world decision-making, as discussed in Section 2.1. One proposal given in the literature is to use the mean squared error as a way of encompassing both bias and variance.

Potentially different criteria will be needed for different outcomes, e.g., when considering primary and secondary outcomes, which may then lead to using different estimators for different outcomes. As well, in some trial settings such as multi-arm trials (and drop-the-loser designs) where more than one arm reaches the final stage, the bias of each arm could be considered separately, but there may also be interest in calculating e.g. the average bias at the across all arms that are selected. In any case, once criteria for assessing estimators have been decided, the next step is to find potential estimators that can be used for the trial design in question. Part I of this paper series is a starting point to find relevant methodological literature and code for implementation.

For more commonly used adaptive designs, a review of the literature may be sufficient to compare the bias and variance of different estimators. Otherwise, we would recommend conducting simulations to explore the bias and variance of potential estimators given the adaptive trial design. In either case, we recommend assessing the estimators across a range of plausible parameter values and design scenarios, taking into account important factors such as the probability of early stopping or reaching the final stage of a trial. More generally, any simulations should follow the relevant FDA guidelines regarding simulation reports for adaptive designs[42(pp28-29)], see also guidance by Mayer et al.[103].

The simulation-based approach can also be used when there are no proposed alternatives to the standard MLE for the trial design under consideration. Even in this setting, we would still encourage an exploration of the bias properties of the MLE. If there is a potential bias of a non-negligible magnitude, then this can impact how the results of the trial are reported (see Section 5.4).

## 5.2 Pre-specification of analyses

The statistical analysis plan (SAP) and health economic analysis plan (HEAP) should include a description of the estimators that are planned to be used to estimate treatment effects of interest when reporting the results of the trial, and a justification of the choice of estimators based on the investigations conducted during the planning stage. This reflects the FDA

guidance[42(p28)], which states that there should be "prespecification of the statistical methods that will be used to […] estimate treatment effects…" and "evaluation and discussion of the design… which should typically include [...] bias of treatment effect estimates". The trial statistician and health economist should work together to develop plans that are complementary to both their analyses.

When available, unbiased or bias-reduced estimators should be used and (in line with the ACE guidance[3,4]) reported alongside the standard MLE. In settings where multiple adjusted estimators are available and are of interest, one adjusted estimator should be designated the 'primary' adjusted estimator for the final reporting of results, with the others included as sensitivity or supplementary analyses (depending on the estimand of interest[104,105]). This is to aid clarity in the interpretation of the trial results, and to avoid 'cherry-picking' the estimator that gives the most favourable treatment effect estimate. Similarly, when only one adjusted estimate is reported alongside the standard MLE, it should be made clear which one is the 'primary' result. More generally, guidelines for adaptive designs should have a clear requirement to consider bias and bias-adjustment when analysing trial results.

As an example of what this looks like in practice, we point the reader to the TAO (Treatment of Acute Coronary Syndromes with Otamixaban) trial as described by Steg et al.[106], particularly their Appendix B, Section 9. The authors consider the MLE and a median-unbiased estimator (MUE), and explore the bias and MSE via simulations and conclude that the MUE has a "consistently smaller" bias with no "noticeable difference in terms of MSE". Therefore, they propose to use the MUE as the point estimator in their trial.

We have deliberately avoided making recommendations on the most appropriate adjustment method because the most appropriate choice of estimator depends heavily on the context and goals of the trial, as well as the type of adaptive design (and trial adaptations) in question. In addition, given that estimation for adaptive designs is an ongoing research area, there is a risk that any recommendations may become outdated. However, for some adaptive designs, such as the group sequential design presented in our case study in Section 4, it is possible to provide stronger guidance (see the discussion in Section 4.1 as an example).

## 5.3 Data Monitoring Committees (DMCs)

When presenting interim results to DMCs, the issue of potential bias should also be considered. We would recommend that the sensitivity of the standard MLE (based on the interim data) to potential bias should be reported, for example based on simulations conducted during the planning stage. As recommended by Zhang et al.[100] and Shimura et al.[90], when unbiased or bias-reduced estimators are available, these should also be presented to the DMC, as an additional tool for appropriately considering potential bias in the decision-making process of whether to stop a trial early (or to perform other trial adaptations such as modifying the sample size).

## 5.4 Reporting results for a completed trial

When reporting results following an adaptive design, there should be a clear description of the "statistical methods used to estimate measures of treatment effects"[3(p16)]. Hence, when unbiased or bias-adjusted estimators are used, this should be made clear, along with any underlying assumptions made to calculate them (for example, being unbiased conditional on the observed stopping time). As reflected in the ACE guidance[3(p16)], "when conventional or naive estimators derived from fixed design methods are used, it should be clearly stated" as well.

The FDA guidance on adaptive designs[42(p30)] states that "treatment effect estimates should adequately take the design into account". Hence, we reiterate that adjusted estimates taking the trial design into account are to be preferred, if available. The FDA guidance goes on to say that "if naive estimates such as unadjusted sample means are used, the extent of bias should be evaluated, and estimates should be presented with appropriate cautions regarding their interpretation"[42(p30)]. Similarly, the ACE guidelines encourage researchers to discuss "Potential bias and imprecision of the treatment effects if naive estimation methods were used".

These discussions would naturally link back to the planning stage literature review and/or simulations (which could potentially be updated in light of the trial results and any unplanned adaptations that took place), taking into account important factors such as the probability of early stopping and plausible values of the unknown true treatment effect. For example, if the potential bias of the MLE is likely to be negligible, this would be a reassuring statement to make. On the other hand, in a setting where no adjusted estimators currently exist in the literature (e.g., for trials which combine multiple trial adaptations together) and there is the potential for non-negligible bias in the MLE, a statement flagging up this potential concern would allow appropriate caution to be taken when using the point estimate to inform clinical or policy decisions, future studies or meta-analyses.

As discussed in Section 5.4, it should be specified in advance (i.e., in the SAP for a confirmatory study) which estimator will be used for the primary analysis and which (if any) estimator(s) will be used as a sensitivity analysis. If an unbiased or bias-adjusted estimator is reported, then it is useful to look at the similarity with the MLE as part of the sensitivity analysis. If the two estimates are very close to each other, then it is reassuring that the trial results appear to be somewhat robust to the estimation strategy used. Conversely, if the two estimates are substantially different, then this may indicate that more care is needed in interpreting the trial results and when using the point estimates for decision-making or further research. However, we caution that the observed difference between the MLE and an unbiased (or bias-adjusted) estimate is not necessarily a precise measure of the actual bias in the observed MLE. Firstly, the bias of the MLE depends on the true underlying treatment effect, which is unknown. Secondly, an unbiased estimator is only unbiased on average, and not necessarily in any particular trial realisation. To this end, a potentially more transparent way of reporting results is to show the plausible extent of bias of the MLE across a practically reasonable range of the true treatment effect, in addition to considering the corresponding probabilities of stopping (or reaching the final stage). Again this would build on the planning stage review and simulations.

Finally, the reporting of appropriate measures of uncertainty for the estimators, such as confidence or credible intervals, is also important. If methods exist for constructing confidence intervals associated with the adjusted estimator, then clearly these can be reported. However, for many adjusted estimators it is not clear how to construct valid confidence intervals, and hence the 'standard' confidence interval for the MLE may be the only one available.

# 6. Discussion

In this paper, we have critically assessed how bias can affect standard estimators of treatment effects in adaptive designs and the negative effects this can have. As discussed in part I of this paper series, there is a growing body of methodological literature proposing unbiased and bias-adjusted estimators for a wide variety of adaptive trial designs. However, as shown in this paper, there has been little uptake of adjusted estimators in practice, with the vast majority of trials continuing to only report the MLE (if indeed it is made clear which estimation method is being used at all). There are a variety of reasons why this may be the case. Firstly, there is a common belief that the bias of the MLE will typically be negligible in realistic trial scenarios. This assumption is sometimes made without supporting evidence such as simulation studies for a variety of trial contexts. In theory, the bias can be very large in certain scenarios[107]. However, as discussed throughout this paper, the magnitude of the bias depends on the type of design and whether we are interested in conditional or unconditional bias. Hence, this issue needs to be carefully considered for the specific trial in question. Secondly, there is perhaps a lack of awareness of the range of different unbiased and bias-adjusted estimators that exist in the methodological literature. Linked with this, statistical software and code to easily calculate adjusted estimators is relatively sparse (see also Grayling and Wheeler[108]), which is an obstacle to the uptake of methods in practice even if they exist. It also remains the case that for more complex or novel adaptive designs, adjusted estimators may not exist (see part I of this paper series, particularly Section 6).

It is our hope that this paper series will encourage the increased use and reporting of adjusted estimators in practice for trial settings where these are available. As described in our guidance section, estimation issues should be considered in the design stage of an adaptive trial. Bowden and Wason[109] give a good example of how this can be done in a principled way for two-stage trials with a binary endpoint. More generally, the estimation strategy should take the design of the trial into account, which motivates the use of adjusted estimators. In terms of trial reporting, statements about the potential bias of the reported estimates can indicate where more care is needed in interpretation of the results and the use of the point estimates for further research including evidence synthesis and health economic analyses.

Finally, to improve the uptake of unbiased and bias-adjusted estimators in practice, there is the need for the further development of user-friendly software and code to allow straightforward calculation of trial results and to aid in simulations. Ideally, the calculation of adjusted estimators could be added to existing widely-used software for adaptive trial design and analysis. Otherwise, there is scope for stand-alone software packages or code (such as that provided for our case study) focusing on estimation after adaptive designs, particularly with simulation studies in mind.

## Acknowledgements

This research was supported by the NIHR Cambridge Biomedical Research Centre (BRC-1215-20014). This report is independent research supported by the National Institute for Health Research (Prof Jaki's Senior Research Fellowship, NIHR-SRF-2015-08-001). The views expressed in this publication are those of the authors and not necessarily those of the NHS, the National Institute for Health Research or the Department of Health and Social Care (DHCS). T Jaki and DS Robertson received funding from the UK Medical Research Council (MC_UU_00002/14). DS Robertson also received funding from the Biometrika Trust. B Choodari-Oskooei was supported by the MRC grant (MC_UU_00004_09 and MC_UU_12023_29). The Centre for Trials Research receives infrastructure funding from Health and Care Research Wales and Cancer Research UK.

## Data Availability Statement

All of the data that support the findings of this study are available within the paper and supplementary information. For the purpose of open access, the author has applied a Creative Commons Attribution (CC BY) licence to any Author Accepted Manuscript version arising.

# Appendix

## A.1 Guidance on bias-adjusted analyses for adaptive designs

**FDA: Adaptive Designs for Clinical Trials of Drugs and Biologics**

"Adaptive designs require specific analytical methods to avoid increasing the chance of erroneous conclusions and introducing bias in estimates. For complex adaptive designs, such methods may not be readily available, and simulations are often critical" - page 6

"It is important that clinical trials produce sufficiently reliable treatment effect estimates to facilitate an evaluation of benefit-risk and to appropriately label new drugs, enabling the practice of evidence-based medicine. Some adaptive design features can lead to statistical bias in the estimation of treatment effects and related quantities. For example, each of the two cases of Type I error probability inflation mentioned in section III.A. above has a potential for biased estimates. Specifically, a conventional end-of-trial treatment effect estimate such as a sample mean that does not take the adaptations into account would tend to overestimate the true population treatment effect. This is true not only for the primary endpoint which formed the basis of the adaptations, but also for secondary endpoints correlated with the primary endpoint. Furthermore, confidence intervals for the primary and secondary endpoints may not have correct coverage probabilities for the true treatment effects.

For some designs there are known methods for adjusting estimates to reduce or remove bias associated with adaptations and to improve performance on measures such as the mean squared error (e.g., Jennison and Turnbull 1999; Wassmer and Brannath 2016). Such methods should be prospectively planned and used for reporting results when they are available. Biased estimation in adaptive design is currently a less well-studied phenomenon than Type I error probability inflation, however, and methods may not be available for other designs. For these other designs, the extent of bias in estimates should be evaluated, and treatment effect estimates and associated confidence intervals should be presented with appropriate cautions regarding their interpretation." - page 8

"Finally, conventional fixed sample estimates of the treatment effect such as the sample mean tend to be biased toward greater effects than the true value when a group sequential design is used. Similarly, confidence intervals do not have the desired nominal coverage probabilities. Therefore, a variety of methods exist to compute estimates and confidence intervals that appropriately adjust for the group sequential stopping rules (Jennison and Turnbull 1999). To ensure the scientific and statistical credibility of trial results and facilitate important benefit-risk considerations, an approach for calculating estimates and confidence intervals that appropriately accounts for the group sequential design should be prospectively planned and used for reporting results." - pages 12-13

"Consider group sequential designs: It is widely understood that multiple analyses of the primary endpoint can inflate the Type I error probability and lead to biased estimation of treatment effects on that endpoint. Less well appreciated, however, is that Type I error probability inflation and biased estimation can also apply to any endpoint correlated with the primary endpoint (Hung et al. 2007)." - page 22

[Documentation Prior to Conducting an Adaptive Trial] "Evaluation and discussion of the design operating characteristics, which should typically include Type I error probability; power; expected, minimum, and maximum sample size; bias of treatment effect estimates; and coverage of confidence intervals. Such evaluations might be achieved through analytical calculations and/or computer simulations. If operating characteristics are evaluated analytically, appropriate details (e.g., literature references or proofs) for the methodology should be submitted." - page 28

"Appropriate reporting of the adaptive design and trial results … For example, the trial summary should describe the adaptive design utilized. In addition, treatment effect estimates should adequately take the design into account, or if naive estimates such as unadjusted sample means are used, the extent of bias should be evaluated, and estimates should be presented with appropriate cautions regarding their interpretation." - page 30

**The Adaptive designs CONSORT Extension (ACE) statement**

"A goal of every trial is to provide reliable estimates of the treatment effect for assessing benefits and risks to reach correct conclusions. Several statistical issues may arise when using an AD depending on its type and the scope of adaptations, the adaptive decision-making criteria and whether frequentist or Bayesian methods are used to design and analyse the trial. Conventional estimates of treatment effect based on fixed design methods may be unreliable when applied to ADs (for example, may exaggerate the patient benefit). Precision around the estimated treatment effects may be incorrect (for example, the width of confidence intervals may be incorrect). Other methods available to summarise the level of evidence in hypothesis testing (for example, p-values) may give different answers. Some factors and conditions that influence the magnitude of estimation bias have been investigated and there are circumstances when it may not be of concern. Secondary analyses (for example, health economic evaluation) may also be affected if appropriate adjustments are not made. Cameron et al. discuss methodological challenges in performing network meta-analysis when combining evidence from randomised trials with ADs and fixed designs. Statistical methods for estimating the treatment effect and its precision exist for some ADs and implementation tools are being developed. However, these methods are rarely used or reported and the implications are unclear. Debate and research on inference for some ADs with complex adaptations is ongoing. In addition to statistical methods for comparing outcomes between groups (item 12a), we specifically encourage authors to clearly describe statistical methods used to estimate measures of treatment effects with associated uncertainty (for example, confidence or credible intervals) and p-value (when appropriate); referencing relevant literature is sufficient. When conventional or naïve estimators derived from fixed design methods are used, it should be clearly stated. In situations where statistical simulations were used to either explore the extent of bias in estimation of the treatment effects … or operating characteristics, it is good practice to mention this and provide supporting evidence (item 24c)." - page 16

"For AD randomised trials, further discussion should include the implications of: …
- Potential bias and imprecision of the treatment effects if naïve estimation methods were used;" - page 21

"For some AD randomised trials, methods to derive statistical properties analytically may not be available. Thus, it becomes necessary to perform simulations under a wide range of plausible scenarios to investigate the operating characteristics of the design (item 7a), impact on estimation bias (item 12b), and appropriateness and consequences of decision-making criteria and rules. In such cases, we encourage authors to reference accessible material used for this purpose (for example, simulation protocol and report, or published related material). Furthermore, it is good scientific practice to reference software, programs or code used for this task to facilitate reproducible research." - page 24

### A.2 Case study: Group sequential design

**Definition of the information at stages 1 and 2**

At stage $k$ ($k = 1,2$), let $\tilde{p}_k$ denote the pooled estimate of the mean overall success probability, i.e., the total number of observed successes divided by the total number of subjects. Then the observed information $I_k$ is given by

$$I_k = \frac{1}{\tilde{p}_k(1-\tilde{p}_k)(1/n_{0k}+1/n_{CEk})}$$

where $n_{0k}$ and $n_{CEk}$ are the number of subjects on the placebo and CE arms, respectively, at stage $k$.

**Conditional versus unconditional perspectives**

Below we give some quotations from the literature focusing on the issue of the conditional versus unconditional perspective in the context of group sequential designs.

*Troendle and Yu (1999)*[98(pp1617-1618)]:
"Suppose a group sequential clinical trial is undertaken to determine the effect of an experimental drug on the state of a certain disease. Now suppose it is known that the trial was stopped at the first interim analysis because of treatment efficacy, and that the estimated treatment effect was $X_1 - Y_1$, the difference in sample means from the two groups... Is $X_1 - Y_1$ a reasonable estimate of the effect size? Although $X_1 - Y_1$ is unbiased, the general estimator $X_T - Y_T$, where $T$ is the random stopping time, is known to be biased. Recently, an unbiased estimator ... and an essentially unbiased estimator ... have been developed for this problem. However, as will be shown later, these methods remain unbiased by overestimating the effect when there is early stopping while underestimating the effect when the trial stops later. The overall effect is an unbiased estimator, but does that leave the scientist, who knows $T = 1$ any happier? We propose conditioning on the stopping time in a group sequential trial to reduce the discrepancy between the conditional expectation of the estimator and the parameter value."

*Fan et al. (2004)*[99(pp506-507)]:
"We also note that the bias referred to is the marginal or overall bias [i.e. the unconditional bias]. As much as the importance of the marginal bias, sometimes we will also face the question of what the potential bias is given the fact that the study is already stopped at this time, especially when it is a very early interim stage. To answer this question, we feel it is more relevant to investigate the bias conditioning on the actual stopping time. In this paper, we focus on the angle of the conditional bias and in the meanwhile also keep in mind the marginal bias."

"The conditional method is not meant to replace the unconditional methods because these two methods are developed to address different issues. Instead it is proposed as an addition and alternative means that we can take advantage of when the conditional bias is more concerning, rather than a replacement to the unconditional methods."

*Zhang et al. (2012)*[100(p4876)]:
"Although this article focuses on the bias conditional on the observed stopping time, we also recognize the importance of the marginal or unconditional bias … Evaluation of the unconditional bias is particularly helpful in the trial design stage; however, there is also value in assessing the potential bias given that the trial has already stopped (conditional bias), especially on the basis of a very early interim analysis. Fan and colleagues found that the conditional bias may be quite serious, even in situations in which the unconditional bias is acceptable[99]. Most of the available adjustment methods focus on the unconditional bias, which has little effect on the conditional bias.

*Schoenbrot and Wagenmakers (2018)*[101(p140)]:
"Although sequential designs have negligible unconditional bias, it may nevertheless be desirable to provide a principled "correction" for the conditional bias at early terminations, in particular when the effect size of a single study is evaluated."

*Shimura et al. (2017)*[102(p2068)]:
"A reduction in conditional bias is as important as a reduction in overall bias because, in practice, researchers can only obtain an estimate that is conditional on the stopping stages."